# GPU-accelerated modeling of biological regulatory networks

Joyce Reimer
Vaccine and Infectious Disease Organization
University of Saskatchewan
Saskatoon SK, Canada
joyce.reimer@usask.ca

Pranta Saha
Vaccine and Infectious Disease Organization
University of Saskatchewan
Saskatoon SK, Canada
prs527@mail.usask.ca

Chris Chen
Vaccine and Infectious Disease Organization
University of Saskatchewan
Saskatoon SK, Canada
oxn302@mail.usask.ca

Neeraj Dhar
Vaccine and Infectious Disease Organization
University of Saskatchewan
Saskatoon SK, Canada
neeraj.dhar@usask.ca

Brook Byrns
Information & Communications Technology
Advanced Research Computing
University of Saskatchewan
Saskatoon SK, Canada
brook.byrns@usask.ca

Steven Rayan
Centre for Quantum Topology and Its Applications (quanTA)
University of Saskatchewan
Saskatoon SK, Canada
steven.rayan@usask.ca

Gordon Broderick
Vaccine and Infectious Disease Organization
University of Saskatchewan
Saskatoon SK, Canada
awr794@mail.usask.ca

## ABSTRACT

The complex regulatory dynamics of a biological network can be succinctly captured using discrete logic models. Given even sparse time-course data from the system of interest, previous work has shown that global optimization schemes are suitable for proposing logic models that explain the data and make predictions about how the system will behave under varying conditions. Considering the large scale of the parameter search spaces associated with these regulatory systems, performance optimizations on the level of both hardware and software are necessary for making this a practical tool for *in silico* pharmaceutical research. We show here how the implementation of these global optimization algorithms in a GPU-computing environment can accelerate the solution of these parameter search problems considerably. We carry out parameter searches on two model biological regulatory systems that represent almost an order of magnitude scale-up in complexity, and we find the gains in efficiency from GPU to be a 33%–43% improvement compared to multi-thread CPU implementations and a 33%–1866% increase compared to CPU in serial. These improvements make global optimization of logic model identification a far more attractive and feasible method for *in silico* hypothesis generation and design of experiments.

## 1 Introduction

The capacity for data-driven fields such as genomics, proteomics, and metabolomics to process high volumes of biological data is at an all-time high, and has resulted in an abundance of undirected signaling interaction networks [1, 2, 3]. Such networks provide essential clues to the regulatory dynamics at play in biological systems; however, interaction networks alone without regulatory parameters such as decisional weights, update kinetics and control action polarities are insufficient for making predictions about the dynamic behaviour of the system under varying conditions.

An impediment to the further parameterization of these networks lies in the challenges found in capturing dynamic rather than static behavior – the costs of time-course experiments are high and getting an adequate number of time-course samples can be challenging, particularly with large, diverse networks [4]. From the computational side, fitting detailed time-series data using conventional methods of differential equations is an expensive optimization process and is highly context-specific because a model that describes one system's dynamics is not typically transferable to other systems. One solution to this that was proposed by Thomas and d'Ari in 1990 [5] and has gained traction in recent studies [4, 6, 7, 8, 9] has been to use a discrete logical network, e.g. binary or tertiary parameters, pertaining, perhaps, to "On/Off" or "Low/Medium/ High", to constrain the search space to a finite number of parameters. This method has found reasonable success in a broad range of situations, provided that a qualitative solution is sufficient for the goals of the study, which is often the case when the goal is to direct experimental design or assist in generating experimental hypotheses. Still, in the case of protein-protein interaction networks that verge on hundreds or thousands of entities and tens of thousands of interactions, the number of parameters that must be fit to data is high, resulting in a huge parameter search space. Furthermore, due to an often complex fitness landscape rife with local minima, and the increasingly large combantorial interger solution space of this logic framework, classical gradient-based methods alone are inadequate. This complex landscape is better addressed with global search methods such as the far more computationally expensive brute-force methods of genetic algorithms or simulated annealing. In these settings, serial CPU implementations quickly become inefficient for finding the global optima in reasonable time of large biological logic networks.

In this study, we demonstrate the speedup that is possible when a GPU-accelerated simulated annealing algorithm is applied to two different biological networks, representing approximately an order of magnitude difference in size—the hypothalamic-pituitary-gonadal (HPG) axis with 102 model parameters and a *Mycobacterium tuberculosis* (Mtb) host-pathogen interaction network, with 723 iteger ternary logic parameters. We compare various GPU configurations against both a serial and parallel CPU implementation and find that most of the GPU configurations offer significant performance advantages to both networks, though some are more efficient than others. These performance advantages are observed in both the serial and parallel CPU comparisons, and they become more significant as the network size and compelxity is increased.

## 2 Methodology

For a detailed and generalizable presentation of the modeling framework, we refer to [4, 9]. What is presented here is a description of the model parameters, temporal update scheme, objective function, and regulatory networks used in this particular study.

### 2.1 Model Parameters

In the discrete logic formulation of biological regulatory networks, individual biological entities (proteins, signaling molecules, etc.) are referred to as "nodes" and the directed regulatory control actions between two entities as "edges".

State values of nodes can take on one of three activation values at a given time step:
- $s_i \in \{0, 1, 2\}$ is the state value of entity $i$, where 0 is associated with low, 1 with moderate, and 2 with high levels of that entity.

Each interaction (edge) has three parameters associated with it. Where the source node is $j$ and target node is $i$,
- $w_{ji} \in \{0, 1, 2\}$ is the weight of edge $ji$, where 0, 1, and 2 denote non-existent, moderate or strong interactions respectively.
- $p_{ji} \in \{-1, 0, 1\}$ is the polarity of the edge $ji$, where -1 denotes an inhibitory effect, 1 an activating effect, and 0 a non-existent one.
- $t_{ji} \in \{1, 2\}$ is the threshold that denotes the minimum level of $s_j$ required for the interaction $ji$ to be active.

All these parameters contribute to determining each node's transition target state, or "image", at any given time step, wherein the image $K_i$ of node $i$ is the cumulative input of all source nodes $j \in J_i$ that have $i$ as a target:

$$K_i = \sum_{j \in J_i} s_j // t_{ji} * w_{ji} * p_{ji}$$

The difference between the image $K_i$ and the current value of $s_i$ acts as a "force" on $s_i$, pushing it toward the value of $K_i$. The number of steps that $s_i$ actually moves in that direction is determined by whether $s_i$ is already at the minimum or maximum allowable state, as well as by its regulatory kinetic parameters. This is further outlined in the following section.

### 2.2 Modeling Regulatory Networks Over Time

To model a regulatory system in time, an update scheme must be selected. Typically, either a synchronous or an asynchronous update scheme is employed [4, 5, 6]. In this study, we use the priority update scheme, presented by Lyman et al. in [9]. Rather than update all nodes at once in every time step as with synchronous updating, or randomly choosing the nodes to be updated as with asynchronous updating, priority updating entails assigning a priority to all nodes at each time step, and then only updating the nodes with the highest priority. This approach has been shown to capture more complex dynamics and fit biological data more closely compared to conventional methods [9].

The priority update scheme requires the definition of three additional nodal kinetic parameters:
- $u_i \in \{1, 2, 3, 4, 5, 6\}$ or $\{10, 11, 12\}$ is the speed of action of node $i$, the former range assigned to entities known to be slower and the latter to those known to be faster. These parameters increase or decrease a node's priority to be updated.
- $m_i \in \{0, 1, 2\}$ is the memory parameter of $i$, i.e., how many consecutive time steps the node $i$ must be told to update before transitioning to the prescribed state.
- $d_i \in \{1, 2\}$ is the step size, the maximum number of state levels $i$ can change per time step.

If a node is eligible to update—meaning its image is different from its current state, it has freedom to increment or decrement in the direction of the image, and its memory criteria has been met—then its priority to be updated is computed as the product of its speed $u_i$ and how many time steps it has been since it was last updated. The nodes sharing the highest priority class are then updated at the end of the current time step, moving a maximum of $d_i$ steps toward $K_i$.

### 2.3 Optimization Objective Function

Given a set of model parameters, the framework defined in Sections 2.1 and 2.2 can be used to generate theoretical time-course data of the biological entities in the network evolving over time. When experimental time-series data is available, a useful goal is to determine a set of parameters that generates simulated trajectories that match these exprimental time courses as closely as possible. This can be accomplished by first discretizing the experimental time course into time steps and sorting the corresponding measured biomarker expression values into bins that match the discrete state levels of the modeling framework; for example, "Low" (0), "Moderate" (1), and "High" (2) as we use here. Building off work from [9], we can then pose an optimization problem, wherein the difference between the simulated data and the experimental data is an objective function to be minimized. Furthermore, we choose to weight the element-wise error with the gradient of the experimental data at that time step, to favour solutions delivering smaller errors during periods of high activity. This is formalized as follows.

We define a reference data matrix R as

$$R = \begin{bmatrix} r_{t_0,0} & \cdots & r_{t_0,n} \\ \vdots & \ddots & \vdots \\ r_{t_f,0} & \cdots & r_{t_f,n} \end{bmatrix},$$

where each element $r_{t,i}$ is the discrete value of measured entity $i$ at time step $t \in [t_i, t_f]$, and $n$ is the number of entities in the system. A reference gradient matrix G can then be computed:

$$G = \begin{bmatrix} 0 & 0 & \cdots & 0 \\ (r_{t_1,0} - r_{t_0,0}) & (r_{t_1,1} - r_{t_0,1}) & \cdots & (r_{t_1,n} - r_{t_0,n}) \\ \vdots & \vdots & \ddots & \vdots \\ (r_{t_f,0} - r_{t_f-1,0}) & (r_{t_f,1} - r_{t_f-1,1}) & \cdots & (r_{t_f,n} - r_{t_f-1,n}) \end{bmatrix}.$$

Similarly, the simulated data matrix S can be defined as

$$S = \begin{bmatrix} s_{t_0,0} & \cdots & s_{t_0,n} \\ \vdots & \ddots & \vdots \\ s_{t_f,0} & \cdots & s_{t_f,n} \end{bmatrix},$$

where each element $s_{t,i}$ is the state of node $i$ at time step $t \in [t_i, t_f]$, and $n$ is the number of nodes in the system. We can then define the objective function $y$ as

$$y = \min[(|S(x) - R|) \odot (|G| + 1)],$$

where

$$x = [w, p, t, u, m, d].$$

That is, we aim to find some vector x of model parameters that gives a simulated matrix S with a minimal Manhattan distance from the reference data R, weighted by the gradient matrix G.

## 2.4 Computational Implementation

*2.4.1 Parallel Simulated Annealing Algorithm.* To solve the objective function, a simulated annealing algorithm is used. Simulated annealing is a common method used in global combinatorial optimization problems where the optimization landscape is expected to contain many local minima. Its stochastic nature allows the acceptance, with a certain probability, of worse solutions for the purpose of escaping local basins in favour of a global minimum, while still systematically refining promising solutions as in classical gradient-based methods [10, 11]. In Table I, we outline a parallelized version of simulated annealing, adapted from the synchronous implementation proposed by [12], that we apply to both CPU and GPU computations. All computations use the following parameters: T = 1000, $T_{min}$ = 0.001, $\rho$ = 0.99, and $n_{iterations}$ = 5000. The value of $n_{threads}$ is variable according to the protocol in question; see Results for an outline of all numerical experiments.

*2.4.2 Software Frameworks.* Both CPU and GPU optimization codes are implemented in Python 3.12 with standard Python libraries. The CPMPy constraint solver library is used to ensure the initial solution vectors comply with the network stability constraints outlined at the end of Section 2.5 [13]. The *concurrent.futures* Python module is used for concurrent CPU computations. In the GPU implementation, the Numba Python compiler is used to interface with CUDA and the GPU device [14].

*2.4.3 Hardware Specifications.* All computations are performed on a high-performance cluster compute node. The CPU components utilize Intel Xeon Platinum 8356H processors, and the GPU computations are carried out on Nvidia A100 GPUs.

---

**Require:** $T > T_{min}$; $0 < \rho < 1$; $n_{iterations} > 0$; $n_{threads} > 0$
*# Set scalar optimization parameters*
$\{x_{best}^{(1)}, x_{best}^{(2)}, \ldots, x_{best}^{(n_{threads})}\} = \{x_0^{(1)}, x_0^{(2)}, \ldots, x_0^{(n_{threads})}\}$
*# Initialize each thread's $x_{best}$ with $x_0$, a randomized parameter vector*
**while** $T > T_{min}$ **do**
*# The following is done in parallel with each thread's $x_{best}$:*
　　$\delta_{best}$ = Simulate($x_{best}$)
　　*# Simulate(x) outputs the value of the objective function at x*
　　$\delta_{current} = \delta_{best}$ ; $x_{current} = x_{best}$
　　**for** $n_{iterations}$ **do**
　　　　$x_{candidate}$ = GetCandidate($x_{current}$)
　　　　*# Perturb the current best solution with a random single-element change to get a new candidate solution*
　　　　$\delta_{candidate}$ = Simulate($x_{candidate}$)
　　　　*# Find the candidate solution's error*
　　　　$r \sim \mathcal{U}(0,1)$ ; A = $\exp(\frac{\delta_{current} - \delta_{candidate}}{T})$
　　　　*# Assign r a uniform random number and compute acceptance criterion A*
　　　　**if** $\delta_{candidate} < \delta_{best}$ **or** $r < A$ **then**
　　　　*# Accept the candidate solution if its error is less than $\delta_{best}$, or if the probabilistic event r meets the acceptance criterion*
　　　　　　$x_{current} = x_{candidate}$ ; $\delta_{current} = \delta_{candidate}$
　　　　　　**if** $\delta_{candidate} < \delta_{best}$ **then**
　　　　　　　　$x_{best} = x_{candidate}$ ; $\delta_{best} = \delta_{candidate}$
　　　　　　　　*# Only overwrite $x_{best}$ if the candidate solution is better; otherwise store it in $x_{current}$*
　　　　　　**end if**
　　　　**end if**
　　**end for**
*# The following is done in serial:*
　　$\delta_{best}^{(*)} = \min_i \{\delta_{best}^{(i)}\}$
　　*# Find global minimum error out of all threads*
　　$\{x_{best}^{(1)}, x_{best}^{(2)}, \ldots, x_{best}^{(n_{threads})}\} = x_{best}^{(*)}$
　　*# Use overall best solution to re-initialize all threads*
　　$T = \rho T$
　　*# Reduce temperature*
**end while**

**Table 1: Parallel Simulated Annealing Algorithm**

## 2.5 Construction of Biological Networks and Experimental Data

Two biological regulatory networks are considered as case studies for the model parameterization method in question, the female hypothalamic-pituitary-gonadal (HPG) axis and a host-pathogen network implicated in persistent *Mycobacterium tuberculosis* infection. A schematic of each network's structure is shown in Fig. 1.

*2.5.1 Hypothalamic-pituitary-gonadal Axis.* The HPG axis regulates reproductive hormones, growth, and metabolism in most animals. The human female HPG axis is a canonical example of cyclical biological dynamics in a healthy state, allowing for the assessment of whether the algorithm in question is able to capture complex pulsatile dynamics.

The HPG axis network we utilize here consists of 9 nodes and 25 edges, adapted from [9] to give a network that adheres to the structural constraints outlined in the following subsection.

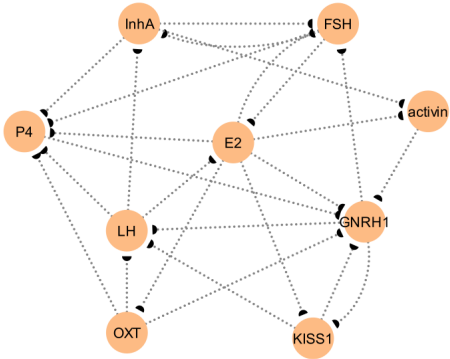

(a) HPG axis network structure

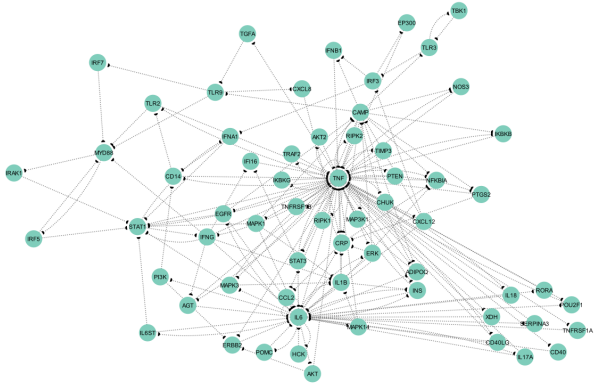

(b) *M. tuberculosis* infection network structure

**Figure 1: Unparameterized structure of the HPG (a) and M. tuberculosis host-pathogen regulatory networks (b).** The unparameterized structures contain nodes and directed edges, but lines are dotted to show that the weight may turn out to be 0 after parameterization, making the edge obsolete. Similarly, target arrows are represented by half circles because the regulatory action may be positive, negative, or 0.

We use the experimental data in [9], collected originally by McLachlan et al. [15], as the reference data to be matched. Of the 9 nodes, experimental data is available for 5—E2, FSH, InhA, LH, and P4. In agreement with [9], we assign FSH, GNRH1, and LH to the high-speed range, and the others to the slower regime (see Section 2.2). Including the initial state, the reference data consists of 20 time points, giving a reference matrix of size 5 by 20.

*2.5.2 M. tuberculosis Host-pathogen Network.* To generate the *M. tuberculosis* host-pathogen network, relevant signalling molecules were autonomously extracted from existing literature using named entity recognition [16], and reading systems from the INDRA environent were used to extract documented relationships from the BEL Large Corpus and the BioPax/Pathway Commons databases [17]. Due to existing literature gaps, the Trustworthy Language Model, a generative AI model that explicitly scores the reliability of responses [18], was used to fill gaps in the network in support of a closed-loop feedback architecture, the characteristics of which are further outlined in the next subsection. The resulting network has 63 nodes and 178 edges. This is the network we use here, representing nearly an order of magnitude scale-up from the HPG network.

Because experimental time course data for this network is not available, we use the *Simulate()* function of our code (Table I) to generate a pseudo-experimental time series as the reference data for this network. To generate this data, the model parameters are randomized within their specified ranges. The constraint satisfaction library CPMPy is used here to ensure the selected weight and polarity parameters adhere to the network constraints outlined in the following subsection. To preserve the ratio of fast to slow entities found in the HPG network, we randomly assign one third of the entities to the fast-node regime. In keeping with the proportion of entities for which data is available in the HPG axis, we only generate reference data for 35 out of the 63 nodes. These 35 nodes are mostly randomly selected; however, we prioritize the selection of nodes that have a more dynamic time course. Data is generated for 20 time points including the initial state, giving a reference matrix of size 35 by 20.

*2.5.3 Auxiliary Constraints.* There are two structural constraints that must be applied to the optimizer to ensure that only sets of network parameters that can logically sustain biological regulation are proposed as solutions.

*a) No sources or sinks:* The first constraint is the network must be a closed-loop system—that is, every node must have both incoming and outgoing interactions with other nodes so that there are no sources or sinks. To enforce this, we apply the following constraint to the weights

$$\sum_{j \in J_i} w_{i,j} \neq 0, \quad \forall i \in I,$$

where, in this case, $J_i$ is a list of all nodes that are targets of node $i$, and $I$ is the set of all nodes. Because all nodes are represented at least once in both the source and target lists, this constraint ensures the existence of both incoming and outgoing edges for every node.

*b) Positive and negative regulation:* The second constraint is that all nodes are required to have an opportunity for both positive and negative regulation, so as to avoid the situation where a node cannot mathematically be activated or inhibited. We ensure that each molecular species will ingerently decay by setting a node bias term of 0; meaning that, if the states of all upstream motivators are at their basal state, each nodes' states can intrinsically decay to 0. The necessary opposing opportunity for positive regulation is enforced by ensuring there is at least one activating upstream relation present for every potential candidate solution vector:

$$\sum_{j \in J_i} p_{i,j_{\{p_{i,j}=1, w_{i,j} \neq 0\}}} \geq 1, \quad \forall i \in I,$$

In this case, $J_i$ is a list of all sources of node i. We sum up all incoming polarities on i that are equal to +1, provided that the weight of that edge is non-zero, and ensure this is at least 1.

Enforcement of these constraints is done at the stage of the *GetCandidate()* function in the simulated annealing algorithm (Table I) by using conditionals when perturbing the weight or polarity components of the parameter vector. We must also consider these constraints when initializing each thread with a solution vector. CPMPy is employed here to efficiently ensure that each thread is given a valid solution vector from which to begin.

## 3 Results

Two sets of seven computations are run to explore the speedup possible when the simulated annealing optimization is run on CPU compared to GPU, with varying numbers of threads. All experiments' configurations are outlined in Table II; these are applied to both the HPG and *M. tuberculosis* networks. We note that all GPU computations are run with a block size of 256, as in [12].

| Device | Number of threads | Number of solution vectors |
|--------|-------------------|---------------------------|
| CPU | 1 | 1 |
| CPU | 32 | 32 |
| GPU | 32 | 32 |
| GPU | 256 | 256 |
| GPU | 1024 | 1024 |
| GPU | 10,240 | 10,240 |
| GPU | 102,400 | 102,400 |

**Table 2: Computational protocol for optimizations.**

### 3.1 Performance of HPG Axis Optimization

The performance profiles of the HPG-axis optimizations are given in Fig. 2A and make evident the overall gains in efficiency when moving from CPU to GPU. The best CPU implementation (32 threads) reaches its best solution ($y$=21) after 87 minutes, and takes almost 200 minutes to reach the stopping conditions. Most of the GPU implementations surpass this, by taking approximately 60 minutes to reach a slightly better answer (70 minutes to reach $y$=20), and finishing the optimization around 150 minutes. This is a moderate performance improvement, representing about a 33% increase in speed when moving from CPU to GPU, as well as the capacity for an improved solution. We note a caveat to this, however, which is that not all GPU implementations result in an improvement; for this problem, employing a very low or very high number of threads can be detrimental to performance. The 32-thread implementation converges quickly, but results in a suboptimal error ($y$=24), and the 102,400-thread configuration takes significantly longer than all other configurations, including those on CPU, to converge (377 minutes to $y$=20; 939 minutes to full convergence).

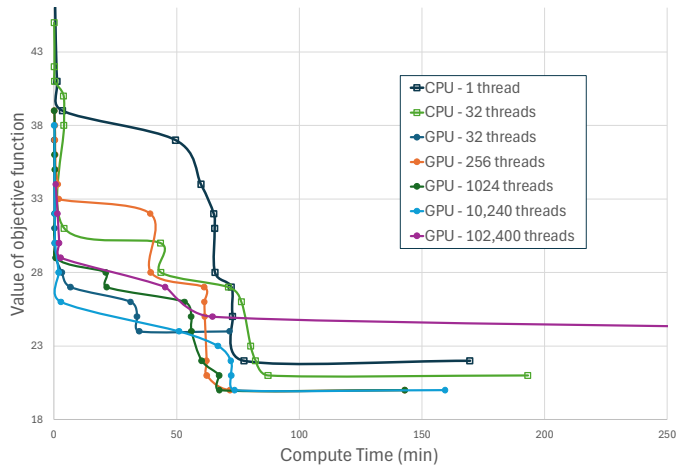

(a) Performance profile of HPG axis optimization.

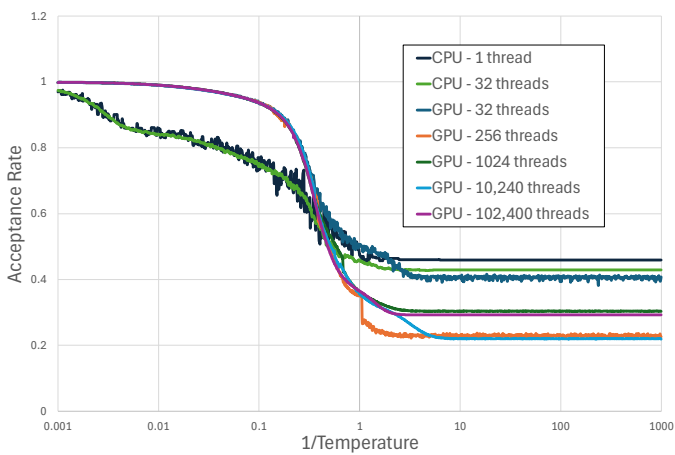

(b) Acceptance rate of simulated annealing algorithm as temperature cools.

**Figure 2: Performance profile (a) and acceptance rate (b) of HPG-network optimization**. a: A plot of each configuration's progress in reducing the objective function over time. The best-performing configurations for this problem seem to be a medium-range of GPU threads, as the 256-, 1024-, and 10,240-thread configurations have the fastest convergence rate and lowest final error. The 102,400-thread configuration takes 939.2 minutes to run; it is left off the graph so as to not distort the other plots, but eventually it converges to the same final error as the other best-performing GPU runs ($y$=20). b: Acceptance rate in the optimization over the progress of the algorithm; i.e., the degree to which worse answers are accepted for the sake of finding even better ones. In this problem, the steeper curves are associated with better performance overall, as is the case with the GPU computations.

In Fig. 2B, the acceptance rates of each optimization are plotted over the course of the simulated annealing algorithm. This time-independent profile is calculated by tracking the number of worse solutions that are temporarily accepted per $n_{iterations}$; i.e. the proportion of iterations in which the acceptance criterion is met by the probabilistic event $r$. This is done for every iteration of temperature reduction in the while-loop (Table I).

The acceptance plots of the CPU implementations have a notably flatter curve than that of the GPU implementations, especially in the range of the algorithm before $T = 1$. Furthermore, the GPU configuration that flattens out first (32 threads) performs the worst in terms of the final objective function value that is achieved. The other four GPU curves, which drop notably lower than the CPU and GPU-32 curves, all (eventually) reach the best function evaluation of $y=20$. Although this curve does not give information about speed to convergence, it serves to demonstrate the role that thorough yet responsive exploration plays in finding accurate solutions to this problem.

The best solution proposed by the 1024-thread optimization is used to produce simulated time-course traces to be compared with the experimental reference data. These plots are shown in Fig. 4. The final objective function evaluation was 20, which is the Manhattan distance weighted by the gradients. The unweighted Manhattan distance between this solution and the reference is 14. Some of them match very closely, for example, LH and P4. The other three are not as close a match, however, the general pulsatile dynamics are still captured by the model.

## 3.2 Performance of *M. tuberculosis* Network Optimization

The *M. tuberculosis* network represents nearly an order of magnitude increase in problem complexity compared to the HPG-axis network. From the performance profiles in Fig. 3A, we can see how the moderate improvements introduced by GPU in Fig. 2A scale up to accommodate a larger problem size. The difference between the speed of GPU and speed of CPU computations is much more evident in these plots; there is a large space between them for much of the computation time, and the CPU computations reach an objective function evaluation of $y=6$ by the end (577 minutes), even though $y=0$ is possible, and is realized by several of the GPU computations.

Unexpectedly, the best GPU configuration for this network is the 32-thread optimization, converging to $y=0$ in just under 400 minutes, and reaching the CPU best solution of $y=6$ in 247 minutes, resulting in an increased speedup of about 43%. The 32-thread configuration performs much better than the moderately-threaded ones that performed best in the HPG network; the 256- and 1024-thread optimizations perform better than CPU for the first 300 minutes, but end up taking about as long as CPU does to converge, and then only converge to $y=6$. The resource-heavy 10,240- and 102,400-thread configurations perform much better than all other configurations for the first two hours, but they then slow down considerably, until finally converging to $y=0$ after 840 and 3516 minutes, respectively. It seems that for this problem, the tradeoff between accuracy and compute time is amplified—except for the outlier case of the 32-thread GPU computation that manages to achieve both.

These performance profiles are further reflected in the acceptance rate plots in Fig. 3B. The configurations that achieve a final evaluation of $y=0$ have a markedly different shape than the ones that only reach $y=6$ (or higher). This is found in the region between T=10 and T=0.1 in particular, where there is a sharp decline in the acceptance rate for the best performing ones. They then jump back up to a mid-range acceptance rate as the solution

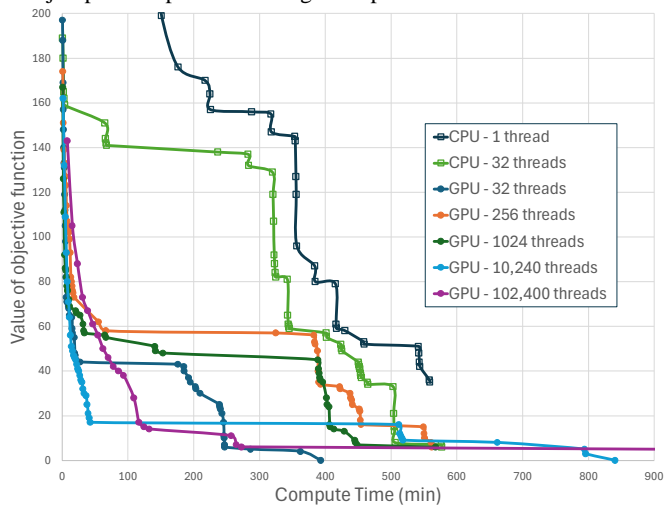

(a) Performance profile of *M. tuberculosis* optimization.

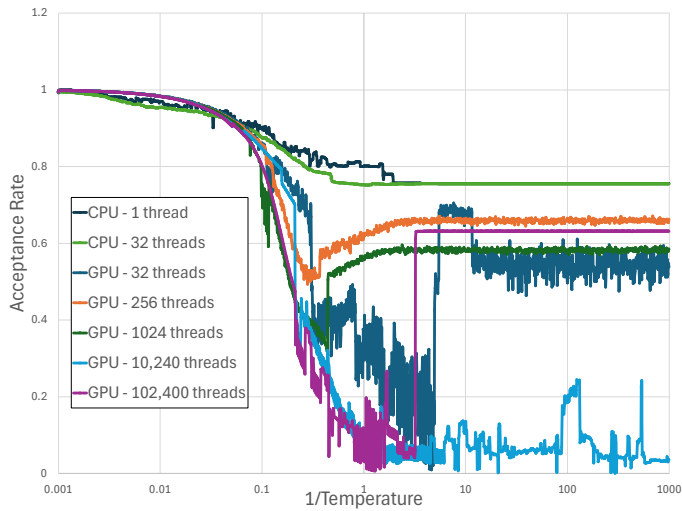

(b) Acceptance rate of simulated annealing algorithm as temperature cools.

**Figure 3: Performance profile (a) and acceptance rate (b) of *M. tuberculosis* network optimization.** a: Each configuration's progress in reducing the objective function over time. The best-performing configurations for this network appear to be a low-medium range of GPU threads; while the higher-threaded configurations perform faster at first, they take much longer than the lower-threaded ones to converge. Again, the 102,400-thread plot is truncated because it takes so long to complete (approximately 7 days), but eventually converges to 0 as well. b: Acceptance rate as the algorithm progresses. Again, steeper curves are generally associated with better performance, seen especially in the difference between the CPU and GPU curves.

tries to converge to 0. It seems that the ones that were the most successful had events of drastic change, either causing a sharp drop in acceptance rate as was the case for the 10,240-threads, or an instantaneous jump up, as was the case for the 32- and 102,400-threads. These may pertain to beneficial stochastic events during the algorithm that are more likely with a higher number of threads, though also seem to happen in the 32-thread case.

The best solution from the 32-thread optimization is used to generate the time-course plots in Fig. 5. It is clear that the simulated data matches the reference data exactly, as is expected for an objective function value of 0.

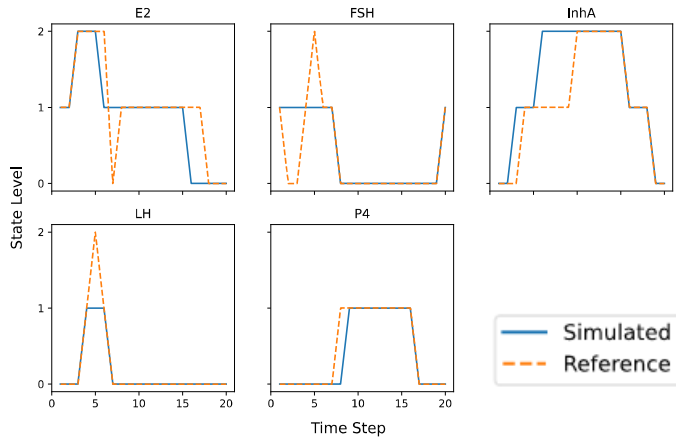

**Figure 4: Simulated time course data compared to reference data, derived from an optimized HPG-axis regulatory network model.** One of the solutions proposed by the optimization with the best objective function evaluation is plotted here, this one from the 1024-thread GPU configuration. It results in an objective function value of 20 (weighted Manhattan distance) and an unweighted Manhattan distance of 14 from the experimental reference data. The nodes with the best fit are LH and P4; however, the other three show a general capturing of the measured activity. Only plots of the 5 nodes for which reference data is available are shown here.

## 4 Discussion

In this work, we have assessed the efficiency gains that are possible when utilizing GPU computing for simulated annealing optimizations in a combinatorial parameter space. To do this, we considered two biologically relevant problems, a smaller and larger regulatory network, representing about an order of magnitude difference in problem complexity, and assessed the increased performance improvement imparted by the GPU architecture as the problem was scaled up.

### 4.1 Performance Improvements

It was shown that simply solving with a GPU improved performance even if using the same number of threads as CPU, as was the case for the 32-thread computations. This can be attributed to the differences in architecture between CPU and GPU—CPUs are designed for general purpose, mainly serial, computing, whereas GPUs are designed to be optimal for massively parallel operations. This problem, both the function evaluation and the optimization itself, involves many iterative loops and repetitive tasks, and therefore is quite well suited for GPU computing. As the number of GPU threads increased, this generally improved the overall performance further. This trend seems to plateau when increasing from 10,240 to 102,400 threads however, especially in the HPG axis network, as this configuration led to extremely long computation times to fully converge. This is likely due to the additional administration work required for the CPU portion of the algorithm to process the latest solution vectors and redistribute the best one before moving on to the next temperature step. When this high-threaded configuration is applied to the larger *M. tuberculosis* network, it seems that efficiency starts to balance out the high overhead; despite still taking a long time to reach $y=0$, it reaches a reasonably low error faster than most of the other configurations. It is likely that with even larger networks, the overhead associated with a higher number of threads will continue to be balanced out by the gains in efficiency.

### 4.2 Acceptance Rates

In general, it appears that a steeper acceptance rate curve is associated with greater success, and sharp drops reflect stochastic events that further improve the outcome of the solution (see 256-threads in Fig. 2B and 32- and 10,240-threads in Fig. 3B). It is notable that CPU acceptance curves look noticeably less steep than GPU acceptance curves, despite the plots being time-independent and thus not fully explained by GPUs' greater capacity for fast parallel processing. In addition to having a more gradual curve, the CPU curves also do not have nearly the stochasticity that the GPU curves have, in that there are no sudden jumps resulting in beneficial probabilistic events.

This can likely also be attributed to the differences in architecture between GPU and CPU, particularly the distinct ways that each handles memory. Random number generation, a key component of this algorithm, is particularly challenging in GPU settings, where access to global memory is limited. Therefore, in a GPU framework, each thread requires its own random number generator that can exist in local memory but that still produces sufficiently distinct numbers from the other threads. The need for a GPU-specific random number generator follows from this, and indeed, there have been many methods proposed for generating random numbers on GPU, the qualities of which tend to vary considerably, depending on the application of interest [19, 20]. In this study, we use Numba's xoroshiro128+ algorithm random number generator for GPU [14]. This particular algorithm seems to introduce relatively more stochasticity to the problem than the CPU's standard Python random module. For a simulated annealing global search algorithm, stochasticity tends to be more beneficial; therefore, it is likely that simply the use of a different random number generator could result in either worse or better performance.

### 4.3 High vs. Low Thread Configurations

A notable outlier in performance profiles is the 32-thread GPU arrangement. In the HPG axis, it starts off performing better than all other GPU computations (aside, perhaps, from the 10,240-thread one), however, around the one-hour mark, it gets surpassed by the other configurations. In the *M. tuberculosis* network, it starts off performing moderately, then quickly exceeds all other configurations. The fact that it gets relatively better than other configurations with a larger problem size suggests that its success may be a product of low overhead. Another factor of its success could be inferred from the sharp drops in the 32-threads' acceptance rate in Fig. 3B between T = 10 and T=1, as this may point to some beneficial stochastic event that quickly propelled it towards a low function value basin.

Two validation tests were run to check whether the success of the 32-thread *M. tuberculosis* optimization was due purely to luck, but both tests showed similar performance: although the final error did not always reach 0 (the maximum error of the tests was $y$=5), these tests converged to their final errors just as quickly as the one plotted in Fig. 3A. This indicates that while the 32 threads' efficiency is not purely stochastic, the achievement of the lowest possible error with 32 threads is.

The success of the 32 threads seems to suggest that, at least for this problem, it is better to spend focused energy refining a small number of solutions than it is to search the parameter space thoroughly with something like ten- or one hundred-thousand initial solutions. Although we can basically guarantee that we will reach the lowest possible error with so many threads (demonstrated in both Fig. 2A and 3A where all high-thread runs reach the lowest error), whether the marginal improvement on error is worth the additional time must be decided on a case-by-case basis. In order to achieve the best balance, a good compromise may be to run several 32-thread GPU computations in parallel with the assumption that each will converge quickly and at least one of them is bound to find the lowest error; however, the drawback is that this will inevitably result in wasted computations.

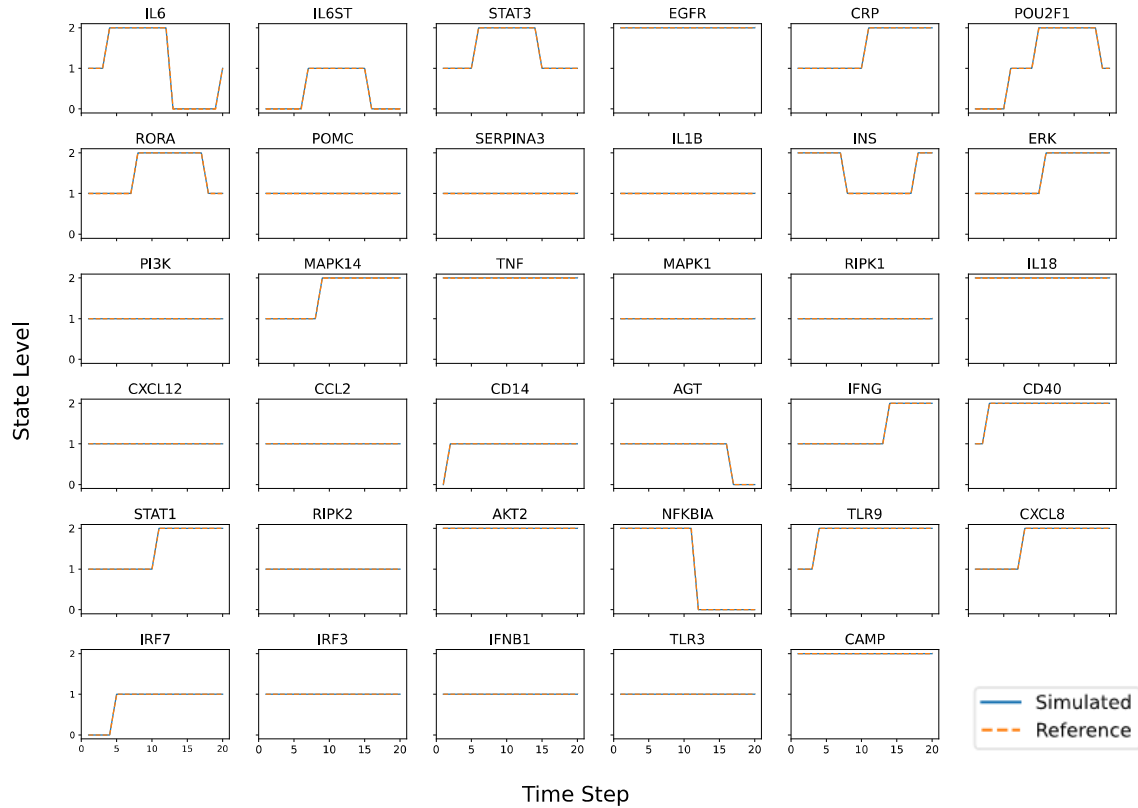

**Figure 5: Simulated time course data compared to reference data, derived from an optimized M. tuberculosis host-pathogen network model.** For the *M. tuberculosis* network, the optimization successfully determined several models that fit the computationally generated reference data perfectly, one of which is used to produce the data here. Only plots of the 35 nodes represented in the reference data are shown.

### 4.4 Comparison with Other Studies

In the present study, we explore the speedup that is possible when GPU computing is applied to a simulated annealing optimization of biological network models. This is not the first study to quantitatively compare CPU to GPU in a simulated annealing context, but it is unique in that it considers both the performance of GPU compared to CPU, as well as the functional practicality of the solutions for a specific experimental application. The optimization involves complex and non-linear function evaluations, which are not often represented in theoretical studies of GPU speedup. These studies tend to use canonical arithmetic functions as benchmarks, as this type of problem showcases GPU speedup more smoothly. Here, we compare our results to a few such studies to identify how our practically-focused implementation compares to more theoretical ones.

A study by Ferreiro et al. compares two different simulated annealing algorithms for GPU with each other (the synchronous algorithm was adapted for use in this study), and with a serial CPU version [12]. Of our experimental setups, the closest configuration to theirs is the 10,240-thread setup. Their change in problem size from 128 to 512 is closest to our increase from the HPG network to the tuberculosis network. What is observed in both their study and ours is a far lower error in GPU than in CPU, even as the problem size is increased. In theirs, the GPU error increases 3 orders of magnitude but remains less than 1 the entire time (for the synchronous implementation), while the CPU error gets close to 10 for the largest problem size. This is similar to what we have seen in this study; in the *M. tuberculosis* network, the error in serial CPU is almost two orders of magnitude higher than that of the high-thread GPU implementations. What is different in our study, however, is the GPU error does not increase at all with a larger problem size, whereas in [12] it does increase, even while staying relatively low. We expect this to indicate that either our implementation or the nature of our problem is such that GPU error will not necessarily increase even if we continue to increase our problem size in future studies.

In [21], a performance comparison is carried out between CPU and GPU in the context of simulated annealing. Their results show that, in general, GPU was mostly faster than CPU, and that the advantage of GPU over CPU increased as problem size was increased. Their computations have 150-400 second execution times, so the scale of the problem is very different from ours, but the relative times do reflect what we see in terms of performance differences—for example, the best GPU compute time in their study is about twice as fast as its CPU counterpart, and we see at least this much speedup in the early stages of the HPG network and in all stages of the *M. tuberculosis* network. It is not clear whether they were using a multi-thread or serial CPU implementation to generate the comparison plot, but either way, similar advantages (two-fold or greater) persist even when comparing against our multi-thread CPU configuration.

The quality of the solutions proposed by the simulated annealing optimization should be assessed as well, especially in the case of the HPG-axis network where a perfect solution is not necessarily guaranteed. Looking at the traces in Fig. 4 in relation to those acquired by Lyman et al. in [9], they are qualitatively quite similar. The LH and E2 traces are about the same level of fitness, our FSH trace appears closer, and their InhA and P4 traces are better. Their solution gives an error of 2.8%, but it is not entirely clear how this was calculated; our best solution is an 8.6% error, which is in the same range. Additionally, it is not specified how long the optimization ran for or what the stopping conditions were, making it harder to directly compare performance.

### 4.5 Limitations and Future Work

The main limitation of the current study is the relatively small networks on which the optimizations were conducted, compared to how large they can be [2, 22, 23]. Although we do see greater performance differentials between GPU and CPU already when moving from 102 to 723 model parameters, it would be ideal to have a wider view of the scaling that happens as we move to networks with several thousand parameters or more. This proof of concept study paves the way for future work to be done on such networks.

There is likely further work that could be done to improve both the current CPU and GPU implementations. On the CPU side, optimizations to the code including rewriting it in a compiled language such as C++ or using Numba as a CPU Python compiler and parallelization framework could improve the performance. On the GPU side, there may be better block or grid sizes, or other thread configurations that would perform better, such as running blocks of 32 threads at a time in parallel, and performing the host reduction step within those groups of 32 rather than spending compute time on reducing too many threads' solutions. This is work to explore in the future; for now, the focus of the study was to show how a GPU architecture designed for massively parallel computations can be leveraged to solve biological simulated annealing optimizations more efficiently than classical CPU methods.

## ACKNOWLEDGMENTS

This work was supported by the University of Saskatchewan's Centre for Quantum Topology and Its Applications (quanTA) and by the Vaccine and Infectious Disease Organization (VIDO). VIDO receives operational funding from the Canada Foundation for Innovation (CFI) through the Major Science Initiatives Fund and from the Government of Saskatchewan through Innovation Saskatchewan and the Ministry of Agriculture. The quanTA Centre's high-performance computational work has been advanced through a CFI John R. Evans Leaders Fund grant while access to quantum computing resources through IBM Quantum and PINQ2 has been made possible by a PrairiesCan Regional Innovation Ecosystems (RIE) contract (both awarded to SR). We thank the University of Saskatchewan Advanced Research Computing (ARC) team for their efforts and close collaboration in creating an excellent local environment for this computational work. This article is submitted with the permission of the Director of VIDO.